\documentclass[aps,prd,eqsecnum,amsmath,amssymb,twocolumn,longbibliography,noeprint,10pt,notitlepage,floatfix,groupedaddress]{revtex4-2}
\usepackage[utf8]{inputenc}

\usepackage[english]{babel}
\addto\captionsenglish{}
\usepackage{multirow}
\usepackage{bm}
\usepackage{latexsym,amsmath,amsfonts}
\usepackage{graphicx}
\usepackage{physics}

\usepackage{footnotebackref}

\usepackage[usenames]{xcolor} 
\usepackage{soul}
\usepackage{makecell} 
\usepackage{rotating}
\usepackage{color}
\usepackage{fancyhdr}
\usepackage{ifthen}
\usepackage{slashed}
\usepackage{xspace}
\usepackage{tabularx}
\usepackage{float}
\usepackage{subcaption}
\usepackage{orcidlink}

\usepackage{booktabs}
\usepackage{amsmath}

\usepackage{placeins}
\usepackage{hyperref}   
\hypersetup{
  colorlinks=true,
  linkcolor=blue,
  citecolor=blue,
  urlcolor=blue, 
}

\begin{document}

\title{Prospects of electromagnetic follow-up of neutron star-black hole mergers in the LIGO-India era}

% \author{Yogita Kumari\orcidlink{0009-0009-0921-2281}}
% \email{yogita.kumari@iucaa.in}
% \affiliation{Inter-University Centre for Astronomy and Astrophysics (IUCAA), Pune 411007, India}

% \author{Kanchan Soni\orcidlink{0000-0001-8051-7883}}
% \email{ksoni01@syr.edu}
% \affiliation{Department of Physics, Syracuse University, Crouse Dr, Syracuse, NY 13210, USA}

% \author{Sanjit Mitra\orcidlink{0000-0002-0800-4626}}
% \email{sanjit@iucaa.in}
% \affiliation{Inter-University Centre for Astronomy and Astrophysics (IUCAA), Pune 411007, India}

\author{{Yogita Kumari$^{1}$}\orcidlink{0009-0009-0921-2281}}
\email[yogita.kumari@iucaa.in]{}

\author{{Kanchan Soni$^{2}$}\orcidlink{0000-0001-8051-7883}}
\email[ksoni01@syr.edu]{}

\author{Sanjit Mitra$^1$\orcidlink{0000-0002-0800-4626}}
\email[sanjit@iucaa.in]{}

\affiliation{$^1$Inter University Centre for Astronomy and Astrophysics, Post Bag 4, Ganeshkhind, Pune 411007}
\affiliation{$^2$Department of Physics, Syracuse University, Crouse Dr, Syracuse, NY 13210}

\date{\today}

\begin{abstract}

Neutron star–black hole (NSBH) mergers are promising multimessenger sources, but only a subset is expected to produce detectable electromagnetic (EM) counterparts, depending on the binary mass ratio, black-hole (BH) spin, and neutron-star (NS) equation of state (EoS). We investigate the prospects for detecting kilonova counterparts to NSBH mergers in the LIGO-India era using the Vera C. Rubin Observatory (Rubin). We simulate GW230529-motivated NSBH populations with two effective tidal-deformability ranges, estimate gravitational-wave detections for the LHV (LIGO - Livingston, LIGO - Hanford, Virgo) and LHVA (LHV with the inclusion of LIGO - India at Aundha) detector networks, and model the detectability of the associated kilonova emission with Rubin. We find that adding LIGO-India approximately doubles the number of EM follow-ups, primarily by increasing the duty cycle and reducing the median sky-localization area, and substantially increases the number of such detections at large distances. For fixed-exposure strategies, the EM detection rate saturates at exposure times of a few hundred seconds, reflecting the trade-off between depth and sky coverage. We therefore introduce an event-specific exposure-time optimization strategy based on the expected counterpart brightness and sky-localization area. With this strategy, the expected NSBH GW+EM detection rate is $4.19_{-3.58}^{+9.66}\,\mathrm{yr}^{-1}$ for the optimistic population and $0.79_{-0.68}^{+1.83}\,\mathrm{yr}^{-1}$ for the conservative population with the LHVA network, compared to $1.46_{-1.25}^{+3.37}\,\mathrm{yr}^{-1}$ and $0.27_{-0.23}^{+0.62}\,\mathrm{yr}^{-1}$, respectively, for LHV. The required Rubin follow-up time remains within the expected Target-of-Opportunity allocation. These results show that in the next decade, a few such NSBH follow-ups may take place, with a fraction of them in the redshift range of $\sim 0.1-0.2$, which will not only be useful for measuring the Hubble constant but can also help us in probing the Hubble parameter.

%
%To date, a few hundred compact binary coalescences (CBCs) have been observed, including one binary neutron star (BNS) merger, GW170817, which was followed up across several electromagnetic (EM) bands, marking the start of multimessenger astronomy. Neutron Star-Black Hole (NSBH) mergers can also have an EM counterpart. With this EM counterpart, one can obtain information on the Equation of State (EoS), and it can also serve as a standard siren to determine the Hubble constant. Detecting a few tens of these mergers with EM counterparts will yield a Hubble constant with 2$\%$ precision. Unlike BNS mergers, not all NSBH mergers have an EM counterpart. It depends on the mass ratio, the BH spin, and the NS EoS. Even if the EM counterpart is present, it must be determined whether it can be detected with current EM telescopes. For this work, we considered the Vera C. Rubin Observatory (also known as LSST). In this work, we tried to estimate the plausibility of detecting the EM counterpart of NSBH mergers with the inclusion of LIGO India. Based on the numbers, we determined whether they could serve as a standard siren. 
\end{abstract}

% \keywords{Classical Novae (251) --- Ultraviolet astronomy(1736) --- History of astronomy(1868) --- Interdisciplinary astronomy(804)}

\maketitle

% Introduction
\section{Introduction}
% With the detection of GW170817, a Binary Neutron Star(BNS) merger, the new era of multimessenger astronomy with GW has begun~\cite{GW170817}. CBCs, like BNS mergers, Neutron Star Black Hole (NSBH) mergers are also potential multimessenger sources for GW and EM observations \cite{corsi2024multimessengerastrophysicsblackholes}. With the detection of GW200105 and GW200115 in third observational run (O3) of the LIGO-Virgo-KAGRA (LVK) collaboration \cite{abbott_observation_2021}, the presence of NSBH mergers is already established, but no EM counterpart has been observed so far for these mergers.

The discovery of GW170817 and its accompanying EM counterpart established multimessenger astronomy as a powerful probe to compact-object physics~\cite{GW170817_eos}, nucleosynthesis~\cite{GW170817_nucleosythesis}, and cosmology~\cite{GW170817_hubble}. Since then, the LIGO-Virgo-KAGRA (LVK) collaboration has detected hundreds of compact binary coalescences, including several NSBH mergers~\cite{theligoscientificcollaboration2026gwtc50introductionversion50}. While binary neutron star (BNS) mergers remain the canonical multimessenger sources, NSBH mergers are also promising systems for joint gravitational-wave (GW) and EM observations~\cite{nsbh_mma, Gupta_2023}. The first confident observations of NSBH mergers, GW200105 and GW200115, during the third observing run of the LVK collaboration confirmed the existence of this population~\cite{abbott_observation_2021}. However, no EM counterpart has yet been identified for an NSBH merger.

An EM counterpart from an NSBH merger is possible if the NS is tidally disrupted before it plunges into the BH. This requires the tidal disruption radius or the Roche limit to lie outside the Innermost Stable Circular Orbit (ISCO) radius \cite{lattimer_black-hole-neutron-star_nodate,foucart_black-hole--neutron-star_2012,foucart_remnant_2018}. Whether disruption occurs depends strongly on the binary mass ratio, the magnitude and orientation of the black hole spin, and the neutron-star compactness, which is determined by the underlying EoS~\cite{foucart_black-hole--neutron-star_2012}. Therefore, only a subset of the NSBH parameter space is expected to produce observable electromagnetic emission. When tidal disruption occurs, a fraction of the neutron-star material remains outside the remnant black hole in the form of dynamical ejecta~\cite{Kyutoku_dyanmical, foucart_dynamical_2017} and accretion-disk outflows~\cite{fernandez_landscape_2020}. This neutron-rich material undergoes rapid neutron-capture (r-process) nucleosynthesis~\cite{Fern_ndez_2017}, powering a kilonova that radiates across optical and near-infrared wavelengths \cite{gompertz_multimessenger_2023}. The detection of such counterparts provides information not accessible using GW observations alone, including improved sky localization, host-galaxy identification, and direct redshift measurements. Thus, NSBH mergers are promising multimessenger sources and potential standard sirens for precision cosmology~\cite{vitale_measuring_2018, Gupta_H0_2023}.

% If there is tidal disruption event (TDE), then there is a possibility of getting a kilonova, which is powered by r-process elements. 
% Lightcurves for these kilonova are calculated using radiative transfer simulations (\cite{tanaka_radiative_2013,kawaguchi_models_2016,kawaguchi_diversity_2020-1}). This lightcurve depends on the remnant ejecta mass from tidal disruption, which, in turn, depends on the physical properties of the binary, such as, component masses, spins, and the NS EoS. 

The recently reported event GW230529~\cite{Abac2024} is a particularly interesting observation in this context. The source is consistent with a compact binary containing a low-mass black hole or a compact object within the lower mass gap. Although the gravitational-wave data are consistent with negligible tidal effects~\cite{Abac2024}, several follow-up studies have shown that the system lies close to the boundary between prompt plunge and tidal disruptions~\cite{Chandra:2024ila}. Depending on the EoS and black-hole spin, ejecta masses ranging from nearly zero to $\sim 0.06\, M_{\odot}$ are possible, potentially producing a faint kilonova~\cite{Chandra:2024ila}. Furthermore,~\citet{Zhu:2024cvt} demonstrated that the probability of tidal disruption can vary significantly with the assumed EoS, ranging from approximately $13\%$ for the AP4 EoS to $63\%$ for the DD2 EoS, with the resulting kilonova expected to reach peak magnitudes of $\sim23$–24.

Despite these favorable prospects, no confirmed EM counterpart was identified from GW230529~\cite{Pillas:2025pfc, Ronchini2024GW230529}. A major limitation was due to its detection in the Livingston detector alone, which resulted in a poorly constrained sky-localization area of approximately $24,100~{\rm deg}^2$~\cite{Abac2024}. Consequently, follow-up observations across gamma ray, optical, and infrared bands were unable to probe a substantial fraction of allowed counterpart parameter space, leaving many viable kilonova models unconstrained \cite{Pillas:2025pfc}. GW230529 therefore illustrates that the primary challenge in establishing NSBH mergers as multi-messenger sources may not be the production of EM emission itself, but rather the ability to localize these events sufficiently well for efficient follow-up.  

In the next-generation gravitational wave detector era, detectors like LIGO-India will play a crucial role in enabling EM follow-up observations. LIGO-India is expected to join the global detector network in the coming years, significantly enhancing the network's sky-localization capabilities through its long geographical baseline and improved detector geometry~\cite{Fairhurst:2012tf, saleem_science_2021,Shukla_2024}. These studies have shown that its inclusion can increase the number of sources localized within $20\,\mathrm{deg}^2$ by a factor of $\sim2{-}3$ and reduce localization areas by up to an order of magnitude for many events. Combined with the wide-field, deep-imaging capabilities of the Vera C Rubin Observatory~\cite{andreoni2024rubin2024envisioningvera} these improvements are expected to substantially increase the success rate of EM counterpart identification and strengthen the scientific potential of multimessenger observations of compact binary mergers~\cite{soni_assessing_2024,Pandey_2025}.

% To constrain the Hubble constant with $\sim 2\%$ accuracy, necessary to resolve the Hubble tension, a few tens of BNS mergers are required~\cite{chen_two_2018}; this required number can be smaller for NSBH mergers~\cite{vitale_measuring_2018}. There are some studies~\cite{Feeney} on the prospects of measuring the Hubble constant with NSBH mergers, but a detailed simulation-based study for EM follow-up has not been made so far. In this work, we discuss the prospects of detecting an EM counterpart from NSBH mergers through a joint observation of the ground-based gravitational-wave detector network and the Vera Rubin Observatory. 

Such improvements are also important for precision cosmology. Because GW170817 demonstrated that multimessenger compact-binary observations can be used as standard sirens~\cite{GW170817_hubble}, increasing the number of well-localized GW events with detectable EM counterparts directly improves the prospects for measuring the Hubble constant. A $\sim2\%$ measurement of the Hubble constant ($H_0$) is particularly important for addressing the current Hubble tension~\cite{Freedman_2021}. Previous studies have shown that a few tens of BNS mergers with EM counterparts could reach this level of precision~\cite{chen_two_2018}, while NSBH mergers may require fewer events in favorable regions of parameter space~\cite{vitale_measuring_2018,Gupta_H0_2023,gupta2024characterizinggravitationalwavedetector, Chen_2024}. In this work, we do not perform an $H_0$ inference directly; instead, we estimate the rate at which NSBH events with detectable EM counterparts could be accumulated in the LIGO-India era. We therefore study the prospects for detecting EM counterparts from NSBH mergers through joint observations with the ground-based GW detector network and the Rubin Observatory.

% In this work, we first simulate two NSBH merger populations with different effective tidal deformability ranges: optimistic and conservative, based on \citet{gompertz_multimessenger_2023}. We then estimate the number of GW detection out of these NBSH merger population with the LHV (Livingston, Hanford and Virgo) detector network, and how this number improves with the inclusion of LIGO-India. Finally, we estimate for how many of these GW detections, EM counterpart can be detected with Vera C Rubin (LSST) observatory when we scan the sky in r and i band twice with different exposure times. We also incorporate the expected peak brightness time, as predicted from the models, in the follow-up scheduling process. As detection of EM counterpart depends on the trade-off between sky localization area and apparent magnitude of source, we introduce a strategy to derive an optimal exposure time to maximise the number of successful EM follow-ups.

We begin by constructing two astrophysically motivated NSBH populations that differ in effective tidal-deformability distributions, representing optimistic and conservative scenarios for EM counterpart production~\cite{gompertz_multimessenger_2023}. Using these populations, we estimate the number of GW detections expected with the LHV detector network and investigate how this number changes when LIGO-India is included. For the GW-detected events, we model the associated kilonova emission and evaluate its detectability with the Rubin Observatory. The follow-up analysis accounts for both the temporal evolution of the kilonova lightcurves and the sky-localization information provided by the GW network. Since successful EM identification depends on a balance between observing depth and sky coverage, we further develop an optimized observing strategy that selects exposure times to maximize the overall number of detectable counterparts. Finally, we estimate the number of NSBH detections with successful EM follow-up by Rubin with the inclusion of LIGO India in the ground-based detector network.

%We find that, with the inclusion of LIGO India in the ground-based detector network, the number of NSBH detections with successful EM follow-up by Rubin increases due to improved sky localization of the source, duty cycle and sensitivity of the network.

This paper is organized as follows: in Sec.~\ref{sec:method}, the simulation framework and the methods for GW signal detection and EM follow-up for NSBH events are outlined. In Sec.~\ref{sec:result}, we present our results. We introduce an optimized strategy for follow-up in Sec.~\ref{sec:opt-strat}. We conclude with discussions in Sec.~\ref{sec:conclusion}.

 % Method
 
\begin{table*}[ht!]
    \centering
    \begin{tabular}{lr}
    \hline
    \hline
    \textbf{Source Property} & \textbf{Distribution}  \\
    \hline
   \rule{0pt}{3ex} BH mass ($m_{\mathrm{BH}}$) 
    & Power law $\propto m_{\mathrm{BH}}^{-\alpha}$ 
     \\
    
    \rule{0pt}{3ex} mass ratio $q=\frac{m_{\mathrm{NS}}}{m_{\mathrm{BH}}}$ 
    & gaussian distribution  $\mathcal{N}(q|\mu,\sigma)$
     \\
    
    \rule{0pt}{3ex} BH spin magnitude $\chi_{\mathrm{BH}}$ 
    & Beta distribution $\propto \chi_{\mathrm{BH}}^{\alpha_{\chi}-1}(1-\chi_{\mathrm{BH}})^{\beta_{\chi}-1}$ 
    \\
    
    \rule{0pt}{3ex} Effective tidal deformability $\tilde{\Lambda}$ 
    & Uniform 
     \\
    \hline
    \hline
    \end{tabular}
    \caption{Intrinsic source-parameter distributions adopted for the simulated NSBH population. Black-hole masses follow a power-law distribution, mass ratios a Gaussian distribution, black-hole spin magnitudes a Beta distribution, and effective tidal deformability a uniform distribution. The distribution hyperparameters are drawn from the population-inference analyses of Refs.~\cite{Biscoveanu2022, Abac2024}, with the tidal-deformability prescription taken from Ref.~\cite{gompertz_multimessenger_2023}.}
    % \caption{Probability distributions used to generate the simulated NSBH population. The BH mass is modeled using a power-law distribution, the mass ratio using a Gaussian distribution, the BH spin magnitude using a Beta distribution, and the effective tidal deformability using a uniform distribution. The hyperparameters of these distributions are drawn from the population-inference results of Refs.~\cite{Biscoveanu2022, Abac2024}, while the tidal-deformability prescription follows Ref.~\cite{gompertz_multimessenger_2023}.} 
    \label{table:pop}
\end{table*}

\section{Simulation Framework}\label{sec:method}

\subsection{NSBH Population Models}\label{sec:pop_gen}

As the first step, we generate the NSBH population based on Refs.~\cite{Biscoveanu2022, Abac2024}. The distributions adopted for the BH mass ($M_{\rm BH}$), mass ratio ($q=\frac{M_{\rm NS}}{M_{\rm BH}}$), BH spin magnitude ($\chi_{\rm BH}$), and effective tidal deformability ($\tilde{\Lambda}$) are summarized in Table~\ref{table:pop}. We consider a target subpopulation with BH masses in the range $2.5-4.5$ $M_{\odot}$ and NS masses ($M_{\rm NS}$) in the range $1.2-2$ $M_{\odot}$. Systems in this region of parameter space are more likely to undergo tidal disruption and therefore produce detectable EM counterparts \cite{Abac2024}. The hyperparameters for these distributions are randomly sampled from the posterior distributions obtained from the population analysis, including GW230529, as provided in \citet{Abac2024}. The corresponding predicted merger rate $55^{+127}_{-47}\mathrm{Gpc}^{-3}\mathrm{yr}^{-1}$  for this parameter range is also obtained from the same reference, as there is no update in the specific mass range we are interested in this paper.  

% The NSBH and BNS mergers impose constraints on the tidal deformability of NSs~\cite{Chatziioannou_2020}.
The properties and detectability of EM counterparts from NSBH mergers depend sensitively on the NS tidal deformability, which is determined by the underlying EoS. Since the true distribution of tidal deformabilities in the NS population remains uncertain, we consider two representative populations. The first allows a broad range of effective tidal deformabilities, $\tilde{\Lambda} \in [0,100]$ (Pop 1), following the approach of \citet{gompertz_multimessenger_2023}. The second is restricted to small tidal deformabilities, $\tilde{\Lambda}\in[0,10]$ (Pop 2), representing a population of more compact neutron stars motivated by recent NICER radius measurements~\cite{Gendreau2022NICER}. In both cases, we assume a uniform distribution in $\tilde{\Lambda}$ and require the corresponding neutron-star radii ($R_{\rm NS}$) to lie within the range $9$–$15$ km.

The simulated NSBH mergers are distributed uniformly in comoving volume corresponding to the luminosity distance range $6-2400$ Mpc. This range is motivated by the order-of-magnitude detectability estimates presented in Appendix~\ref{sec:detect}. For each simulated source, the sky location, polarization angle, and binary inclination are sampled isotropically.

The GW signals from simulated mergers are generated using \textsc{SEOBNRv4\_ROM\_NRTidalv2\_NSBH}~\cite{PhysRevD.102.043023}, a frequency-domain effective-one-body waveform model describing the dominant quadrupolar gravitational-wave emission from aligned-spin NSBH binaries. The model incorporates the effects of neutron-star tidal disruption during the late inspiral and merger. It is calibrated for NS masses in the range $[1,3]M_{\odot}$, tidal deformability in the range $[0,5000]$, spin magnitude up to 0.05, BH spin magnitude up to  0.9, and inverse of the definition of mass ratio used in this paper in the range $1-100$.

Using the population model described above, we generate 90,412 simulated NSBH mergers and distribute them uniformly over a one-year observing period.

\subsection{Gravitational-wave Detection Criteria}\label{sec:gw_det}

To estimate the number of mergers detectable by ground-based GW observatories from the simulated population, we consider two detector-network configurations. The three-detector network, denoted by LHV, consists of the Advanced LIGO detectors at Livingston (L) and Hanford (H)~\cite{Aasi_2015}, together with Advanced Virgo (V)~\cite{ Acernese_2015}. The four-detector network, denoted by LHVA, additionally includes the under construction LIGO-India detector at Aundha~(A)~\cite{Iyer2011LIGOIndia,Abbott_A+_2020}. The LIGO detectors, including LIGO-India, are assumed to operate at A+ sensitivity \cite{Abbott_A+_2020,LIGOT1800042v5}, while Virgo is assumed to operate at the Advanced Virgo design sensitivity~\cite{Virgo2009Advanced,VirgoVIR0596A19}. To account for periods during which the detectors are not collecting science-quality data, we assign an independent duty cycle of 70\% to each detector~\cite{lvk_observing_capabilities}.

% we consider a detector network consisting of the three Advanced LIGO observatories (Livingston, Hanford,  and Aundha) as well as Virgo~\cite{GCN_LVK}. The LIGO detectors are assumed to operate at A+ sensitivity~\cite{LIGOT1800042v5}, while Virgo is assumed to operate at the Advanced Virgo design sensitivity~\cite{VirgoVIR0596A19}. Since these detectors do not collect science data all the time, each detector is assigned a 70\% duty cycle~ \cite{lvk_observing_capabilities}.

A simulated source is classified as gravitational-wave detected if it produces an optimal signal-to-noise ratio (SNR) greater than 4 in at least two operating detectors and a network SNR of at least 12~\cite{singer_first_2014}. The network SNR is calculated using only the detectors that are operational at the time of the merger. For each detected event, we generate a sky-localization probability map using \texttt{BAYESTAR}~\cite{singer_first_2014}. 

% A source is classified as detected if it produces an optimal signal-to-noise ratio (SNR) greater than 4 in at least two detectors and achieves a network SNR of at least 12~\cite{singer_first_2014}.  For each detected event, sky-localization maps are generated using \texttt{BAYESTAR}~\cite{singer_first_2014}.

\subsection{Electromagnetic Follow-up Framework}
\label{sec:EM_det}

\subsubsection{NSBH Kilonova Light-curve Modeling}\label{sec:lc}

To determine whether GW-detected events in Sec.~\ref{sec:gw_det} could be followed up by EM telescopes: First, it is determined whether an EM counterpart is present for the GW-detected source by calculating the remnant mass left outside the BH horizon after merger due to tidal disruption of NS using a semi-empirical formula given in \citet{foucart_remnant_2018}, obtained by fitting numerical simulations. The source frame remnant mass $\mathrm{M}_{\mathrm{rem}}$ for NSBH with NS mass $M_{\mathrm{NS}}$, BH mass $M_{\mathrm{BH}}$, dimensionless BH aligned spin magnitude $\chi_{\mathrm{BH}}$ and NS radius $R_{\mathrm{NS}}$ is given by,
%
% To access the prospects for electromagnetic follow-up of the GW detected from NSBH mergers, we first determine whether tidal disruption of the neutron star leaves sufficient matter outside the black hole horizon. We estimate the amount of this remnant matter using the empirical formula of~\citep{foucart_remnant_2018} calibrated to numerical-relativity simulations of NSBH merger. For a NSBH with NS mass $M_{\mathrm{NS}}$, BH mass $M_{\mathrm{BH}}$, BH spin $\chi_{\mathrm{BH}}$ and NS radius $R_{\mathrm{NS}}$
%
\begin{equation}
   \mathrm{M}_{\mathrm{rem}}=M_{\mathrm{NS}}^b\left[\mathrm{Max}\left(\alpha\frac{1-2C_{\mathrm{NS}}}{\eta^{1/3}}-\beta\hat{\mathrm{R}}_{\mathrm{ISCO}}\frac{C_{\mathrm{NS}}}{\eta}+\gamma,0\right)\right]^{\delta}.
\end{equation}
Here, $\alpha,\beta,\gamma,\delta$ are parameters, given by $\alpha = 0.406, \beta = 0.139, \gamma = 0.255,$ and $\delta = 1.761$ as provided in \citet{gompertz_multimessenger_2023}, $\eta=q/(1+q)^2$ is the symmetric mass ratio, $C_{\mathrm{NS}}=\frac{GM_{\mathrm{NS}}}{R_{\mathrm{NS}}c^2}$ is compactness, $M_{\mathrm{NS}}^b=M_{\mathrm{NS}}\left(1+\frac{0.6C_{\mathrm{NS}}}{1-0.5C_{\mathrm{NS}}}\right)$ is baryonic NS mass, and $\hat{\mathrm{R}}_{\mathrm{ISCO}}={\mathrm{R}}_{\mathrm{ISCO}}/\mathrm{M}_{\mathrm{BH}}$ is the normalized ISCO radius, can be expressed as,
\begin{equation}
    \hat{\mathrm{R}}_{\mathrm{ISCO}}=3+ Z_2- \mathrm{sgn}({\chi}_{\mathrm{BH}})\sqrt{(3-Z_1)(3+Z_1+2Z_2)},
\end{equation}
with $Z_1=1+(1-{\chi}_{\mathrm{BH}}^2)^{1/3} [(1+{\chi}_{\mathrm{BH}})^{1/3}+(1-{\chi}_{\mathrm{BH}})^{1/3}]$ and  $Z_2=\sqrt{3{\chi}_{\mathrm{BH}}^2+Z_1^2}$.
If the remnant mass is zero, there will be no EM counterpart. The effects of precession, NS spin, and magnetic field are neglected in order to calculate the remnant mass here.

For systems with non-zero remnant mass, we generate kilonova lightcurves using a semi-analytic model developed in~\citet{gompertz_multimessenger_2023}. It assumes that there are two types of ejecta present: (i) dynamical ejecta and (ii) outflow from the disk due to thermal and magnetic winds. All ejecta are taken as conical sections parameterized by their half-opening angles. A schematic diagram of the ejecta is shown in Fig.~1 of \citet{gompertz_multimessenger_2023}.
% In this framework, the rem ant material can be bound or unbound from BH based on its velocity relative to the escape velocity\cite{1974ApJ...192L.145L}. There are t o possibilities:(i) Dynamical ejecta, which is unbound material when the merger is going on, (ii) Ou flow from disc, formed around the remnant BH and taken to be aligned with the BH spin, due to thermal processes like  iscosity and nuclear recombination or magnetic field. This dynamica  ejecta mass can be given by a fitting function. It is assumed to be axially symmetric and confined to lower half-opening angles. The disc mass is then just the total remnant mass minus the dynamical ejecta mass. The thermal wind mass is a function of the mass ratio. There are two comp nents of this thermal wind: the blue component (with decreased opacity as the temperature drops over time) enveloping the red component. Mass of Magnetic winds is taken as a fraction of thermal wind. All ejecta are taken as a conical section with half-opening angles. A schematic diagram of ejecta is shown in FIG 1, taken from (\cite{gompertz_multimessenger_2023}).

Following~\citet{gompertz_multimessenger_2023}, we assume that the mass of the magnetically driven wind is uniformly distributed between 0.5 and 1 times the thermally driven ejecta mass. We parameterize the ejecta geometry using the cosine of the half-opening angle. The dynamical ejecta is confined to equatorial regions with $\cos \theta_{dyn} \in [0,0.342]$, while the magnetically driven wind occupies polar regions with $\cos \theta_{mag} \in [0.5, 1]$. The thermally driven wind is distributed within the intermediate angular region. The observer's viewing angle equals the binary inclination angle. 

Using the ejecta masses, velocities, opacities, and angular distributions, we generate kilonova lightcurves following the semi-analytic prescription of~\citet{gompertz_multimessenger_2023}. We first compute the bolometric luminosity of each ejecta component using the model's radiative transfer formalism. Assuming blackbody emission, we then calculate the corresponding spectral energy distribution (SED) for each component. The total observed emission is obtained by projecting the individual ejecta components onto the observer's line of sight and summing their contributions. We neglect interactions between different ejecta components and assume that, in regions where two components overlap, the observed flux is equally shared between them. Throughout this work, we do not include the possibility of a relativistic jet or any associated afterglow emission and restrict our analysis to kilonova emission arising from the dynamical ejecta and disk-driven winds.
% Half opening cosine angle for dynamical ejecta is taken to lie between [0,0.342] and magnetically driven wins between [0.5,1] as in \cite{gompertz_multimessenger_2023}. In between, there will be thermal wind. The viewing angle equals the source's inclination. To get the lightcurve from ejecta masses: First, the bolometric luminosity for each ejecta component is calculated using semi-analytical formulas. Then, with the assumption of blackbody radiation, the Spectral Energy Density (SED) is calculated for each ejecta. At last, the final lightcurve is calculated by projecting the component onto the line of sight. It is assumed that no interaction takes place between different components of ejecta. At the intersection, a 50:50 contribution is taken. 
% The possibility of the existence of a relativistic jet at the pole is ignored.

% \begin{figure}[!t]
%     \centering
%     \includegraphics[width=0.8\linewidth]{ejecta.png}
%     \caption{This schematic diagram, taken from Reference \cite{gompertz_multimessenger_2023} of ejecta from NSBH merger. There are two kinds of ejecta: (i)dynamical ejecta $M_{\mathrm{dyn}}$ due to tidal disruption and (ii) post-merger ejecta due to outflow from the disk formed. Post merger ejecta can be thermally ($M_{\mathrm{th}},v_{\mathrm{th}})$ or magnetically ($M_{\mathrm{mag}},v_{\mathrm{mag}}$) driven. Thermally driven wind has two components, blue and red, due to changes in opacity as the temperature drops. All components are assumed to be indepe dent.}
%     \label{ejecta}
% \end{figure}

\subsubsection{Vera C. Rubin Observatory Detectability Criteria}\label{sec:lsst}

To forecast the detectability of kilonova counterparts with the Rubin Observatory, we use the observation planning toolkit {\tt GWEMOPT}~\cite{coughlin_optimizing_2018}. For each detected NSBH event, we provide the corresponding kilonova lightcurves and GW sky localization map as inputs to \texttt{GWEMOPT}. The localization region is first tessellated using the Multi-Order Coverage (MOC) tiling scheme, which accounts for the telescope field of view. Observation time is then allocated to the resulting tiles based on the sky-localization probability and telescope configuration parameters, including the exposure time and the required number of exposures per tile. Tiles with probabilities less than 1\% of the maximum tile probability are discarded, as are tiles that cannot be observed for a sufficient duration within the available observing window. The remaining tiles are assigned equal exposure times and scheduled using the greedy-slew algorithm implemented in \texttt{GWEMOPT}, which prioritizes tiles with the highest localization probability while accounting for their visibility for the required number of exposures and the telescope's observing constraints.

% First, it tiles the sky based on the telescope's configuration, such as the Field of View. For our purposes, we used MOC (Multi-Order Coverage) tiling. After tiling, time is allocated to each tile based on the skymap and telescope configuration information, like exposure time and the number of exposures required for each tile. Tiles with a low probability (0.01 of the maximum probability) are rejected. With that, tiles that can't be observed for long, even if they have a high probability, are rejected. All other tiles get the same exposure time. At last, scheduling was done using a greedy slew algorithm that selects the tiles with the highest probability first, based on the time allocated to each tile and its availability in the observation window (which depends on the telescope's location). 

To classify a GW event as detected in EM as well, we adopt a criterion similar to that used in \citet{coughlin_optimizing_2018}.  We first check whether the sky position where the source is located is covered, i.e., that at least one scheduled tile contains the source location.  We then check if the source's apparent magnitude at the time of observation is smaller than the telescope's limiting magnitude for the given exposure time.  Throughout this work, each tile is observed twice in both i and r bands of Rubin.  If the source is in one of the observed tiles and has a magnitude below the limiting magnitude in each band in both observations, we call it a detection.  The limiting magnitude is calculated for the given exposure time and airmass using the code in Refs.\cite{andreoni2024etc}.

We observe a particular event with the Rubin observatory for 1 day after 1 day of GW detection, as most lightcurves peak within 1-2 days of the merger~ (See Fig.~\ref{fig:lc_non_zero_500}). All tiles are observed in r and i bands twice. As the Rubin Observatory will have around 3\% of its total observation time for Target of Opportunity (ToO), which will be around 50-60 hrs per year \cite{andreoni2024rubin2024envisioningvera}, we also calculate the total time required to follow up on these EM candidates (including the ones that do not fulfill the above detection criteria, so that we get a realistic estimate of the total ToO time). Net time spent at each tile is calculated as the sum of readout time, slew time, and exposure time. Readout time is taken as 2 sec for each 15 sec exposure and slew time as 4.8 sec~\cite{RubinKeyNumbers2026}.

% With different exposure times, scheduling with the Rubin Observatory is done to determine the optimal exposure time  GW Events are observed after 12 hrs of GW detections, as most lightcurves peak within 1-2 days, for 24 hours or until all tiles are covered in r and i bands twice. 
% We also calculated the total time required to follow up on these EM candidates, as LSST will have around 3\% for Target of Opportunity (ToO), which will be around 50-60 hrs per year \cite{andreoni2024rubin2024envisioningvera}  Time at each tile is taken as the sum of readout time, slew time, and exposure time. Readout time and slew time are taken from Reference \cite{RubinKeyNumbers2026}.

\begin{table*}[ht]
    \centering

    \begin{subtable}{0.48\textwidth}
        \centering
        \begin{tabular}{c|cc|cc}
\hline
\textbf{Exposure} & \multicolumn{2}{c|}{\textbf{Detection rate}} & \multicolumn{2}{c}{\textbf{Total duration}} \\
\textbf{time (sec)} & \multicolumn{2}{c|}{\textbf{(per year)}} & \multicolumn{2}{c}{\textbf{(hours per year)}} \\
\cline{2-5}
     & LHV & LHVA & LHV & LHVA \\
\hline
\rule{0pt}{3ex}30 & $0.89_{-0.76}^{+2.05}$  & $1.82_{-1.56}^{+4.21}$ & $3.27_{-2.80}^{+7.56}$ & $5.10_{-4.36}^{+11.79}$ \\
\rule{0pt}{3ex}60 & $1.02_{-0.87}^{+2.36}$  & $2.49_{-2.13}^{+5.76}$ & $4.86_{-4.15}^{+11.22}$ & $8.61_{-7.36}^{+19.89}$ \\
\rule{0pt}{3ex}120 & $1.05_{-0.89}^{+2.42}$  & $2.98_{-2.55}^{+6.88}$ & $5.91_{-5.05}^{+13.66}$ & $13.63_{-11.65}^{+31.47}$ \\
\rule{0pt}{3ex}180 & $1.11_{-0.95}^{+2.56}$  & $3.22_{-2.76}^{+7.44}$ & $8.48_{-7.25}^{+19.58}$ & $17.37_{-14.84}^{+40.11}$ \\
\rule{0pt}{3ex}300 & $0.99_{-0.84}^{+2.28}$  & $3.24_{-2.77}^{+7.47}$ & $10.61_{-9.07}^{+24.51}$ & $23.26_{-19.87}^{+53.70}$ \\
\rule{0pt}{3ex}600 & $0.64_{-0.55}^{+1.49}$  & $2.24_{-1.91}^{+5.17}$ & $8.87_{-7.58}^{+20.49}$ & $21.95_{-18.76}^{+50.69}$\\
\rule{0pt}{3ex}1200 & $0.18_{-0.16}^{+0.42}$  & $0.86_{-0.74}^{+1.99}$ & $5.34_{-4.56}^{+12.32}$ & $12.20_{-10.42}^{+28.16}$ \\
\rule{0pt}{3ex}1800 & $0.10_{-0.08}^{+0.22}$  & $0.47_{-0.41}^{+1.10}$ & $2.68_{-2.29}^{+6.20}$ & $9.51_{-8.13}^{+21.96}$ \\
\rule{0pt}{3ex}3600 & $0.04_{-0.03}^{+0.08}$  & $0.06_{-0.05}^{+0.14}$ & $2.20_{-1.88}^{+5.08}$ & $6.15_{-5.25}^{+14.19}$ \\[0.15cm]
\hline
\hline
\end{tabular}
        \caption{Pop1}
        \label{table:configure}
    \end{subtable}
    \hfill
    \begin{subtable}{0.48\textwidth}
        \centering
        \begin{tabular}{c|cc|cc}
\hline\hline
\textbf{Exposure} & \multicolumn{2}{c|}{\textbf{Detection rate}} & \multicolumn{2}{c}{\textbf{Total duration}} \\
\textbf{time (sec)} & \multicolumn{2}{c|}{\textbf{(per year)}} & \multicolumn{2}{c}{\textbf{(hours per year)}} \\
\cline{2-5}
     & LHV & LHVA & LHV & LHVA \\
\hline
\rule{0pt}{3ex}30 & $0.13_{-0.11}^{+0.31}$  & $0.30_{-0.26}^{+0.70}$ & $0.97_{-0.83}^{+2.25}$ & $1.73_{-1.48}^{+3.99}$ \\
\rule{0pt}{3ex}60 & $0.16_{-0.14}^{+0.37}$  & $0.40_{-0.34}^{+0.93}$ & $1.42_{-1.22}^{+3.29}$ & $2.50_{-2.14}^{+5.77}$ \\
\rule{0pt}{3ex}120 & $0.19_{-0.17}^{+0.45}$  & $0.54_{-0.46}^{+1.24}$ & $1.76_{-1.50}^{+4.06}$ & $3.94_{-3.37}^{+9.11}$ \\
\rule{0pt}{3ex}180 & $0.21_{-0.18}^{+0.48}$  & $0.58_{-0.50}^{+1.35}$ & $2.35_{-2.01}^{+5.43}$ & $4.79_{-4.09}^{+11.06}$ \\
\rule{0pt}{3ex}300 & $0.19_{-0.17}^{+0.45}$  & $0.57_{-0.49}^{+1.32}$ & $3.21_{-2.74}^{+7.41}$ & $5.67_{-4.84}^{+13.09}$ \\
\rule{0pt}{3ex}600 & $0.13_{-0.11}^{+0.31}$  & $0.47_{-0.41}^{+1.10}$ & $2.95_{-2.52}^{+6.80}$ & $5.15_{-4.40}^{+11.89}$\\
\rule{0pt}{3ex}1200 & $-$  & $0.19_{-0.17}^{+0.45}$ & $1.49_{-1.27}^{+3.43}$ & $3.66_{-3.13}^{+8.45}$ \\
\rule{0pt}{3ex}1800 & $-$  & $0.07_{-0.06}^{+0.17}$ & $1.53_{-1.30}^{+3.52}$ & $3.28_{-2.80}^{+7.57}$ \\
\rule{0pt}{3ex}3600 & $-$  & $0.02_{-0.02}^{+0.06}$ & $0.73_{-0.63}^{+1.69}$ & $1.30_{-1.11}^{+3.00}$ \\[0.15cm]
\hline\hline
\end{tabular}
        
    \caption{Pop2}
    \label{table:configure2}
    \end{subtable}
    \caption{EM counterpart detection rate based on merger rate density ($55^{+127}_{-47}\mathrm{Gpc}^{-3}\mathrm{yr}^{-1}$ as in Ref.~\cite{Abac2024}) are  shown for different exposure times to get optimal detection strategy for Pop1 and Pop2 (See Sec.~\ref{sec:pop_gen}). Each tile is observed twice in and r, and i band with a 24-hour observation start time. The total duration indicates the average time spent over the course of one year following up on EM candidates. In the case of LHV, the EM candidates are lower, so the time taken is also smaller. for Pop2 with LHV network, the number of EM candidates is very small; we are getting null detections for higher exposure time. This can be a finite number that increases with the number of injections, but it will still be very small.}
    \label{tab:conf}
\end{table*}

\section{Results}\label{sec:result}

\subsection{Gravitational-wave detections and EM candidates}\label{sec:results_gw}

\begin{figure}[ht!]
    \centering
    \includegraphics[width=0.96\linewidth]{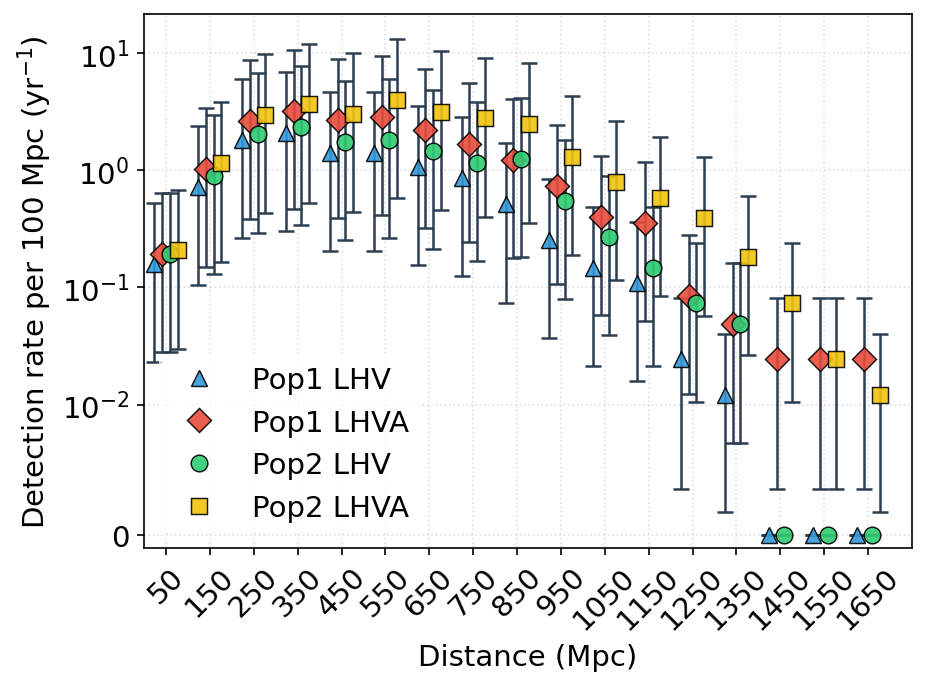}
    \caption{The expected number of NSBH mergers for the LHV and LHVA networks per 100 Mpc distance for Pop1 and Pop2 (See Sec.~\ref{sec:pop_gen}). For Pop1, the median distance is 442.06 and  497.97 Mpc for LHV and LHVA, respectively, and 10\% detections are beyond 801.78 and  878.78 Mpc distance, respectively. For Pop2, the median distance is 487.98 and  561.59 Mpc for LHV and LHVA, respectively, and 10\% detections are beyond 870.67 and  950.98 Mpc, respectively.}
    \label{fig:dist_det}
\end{figure}

We retain only those injections for which the waveform approximant can successfully generate GW signals, as discussed in Sec.~\ref{sec:pop_gen}. This results in 89,943 and 89,916 valid injections for Pop1 and Pop2, respectively. However, when comparing the two populations and quoting detection fractions, we normalize by the total number of injections generated for each population, as specified in Sec.~\ref{sec:pop_gen}. For Pop1, at least two detectors in the LHV network are simultaneously operational for 78.07\% of the total injections. With the inclusion of LIGO-India, this fraction increases to 91.15\% for the LHVA network. Similar duty-cycle fractions are obtained for Pop2, with at least two detectors operational for 78.04\% and 91.12\% of the total injections in the LHV and LHVA networks, respectively. For Pop1, approximately 0.96\% of the total injected sources satisfy our detection criteria in the LHV network, requiring an SNR of $\rho \geq 4$ in at least two detectors and a network SNR of $\rho_{\rm net} \geq 12$. For the LHVA network, the corresponding detection fraction increases to 1.76\%. Thus, the inclusion of LIGO-India increases the number of GW detections by approximately a factor of 1.83, owing to the combined improvements in network duty cycle and sensitivity. For Pop2, the corresponding detection fraction is 1.27\% for the LHV network and increases to 2.42\% for the LHVA network. The inclusion of LIGO-India therefore increases the number of detected sources by a factor of approximately 1.90 for this population.

These detection fractions can be converted into expected numbers of detections per year using the merger rate of $55^{+127}_{-47}\,\mathrm{Gpc}^{-3}\,\mathrm{yr}^{-1}$, reported by \citet{Abac2024}, by multiplying the merger rate by the comoving volume (~$20~\mathrm{Gpc}^3$) and the fraction of detections out of all the injections. The expected number of GW detections as a function of distance for Pop1 and Pop2 is shown in Fig.~\ref{fig:dist_det}.

\begin{figure*}[t]
    \centering

    \begin{subfigure}{0.48\textwidth}
        \centering
        \includegraphics[width=\linewidth]{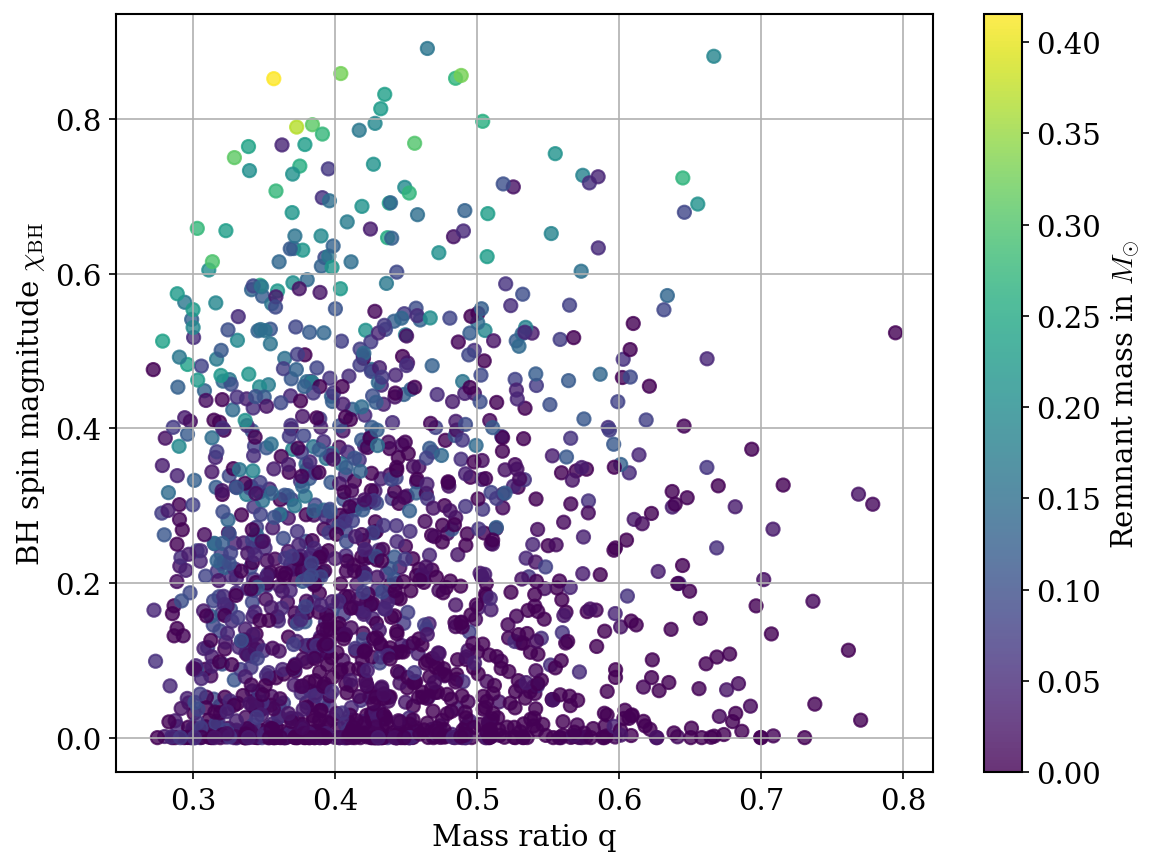}
        \caption{Pop1}
        \label{fig:rem_mass1}
    \end{subfigure}
    \hfill
    \begin{subfigure}{0.48\textwidth}
        \centering
        \includegraphics[width=\linewidth]{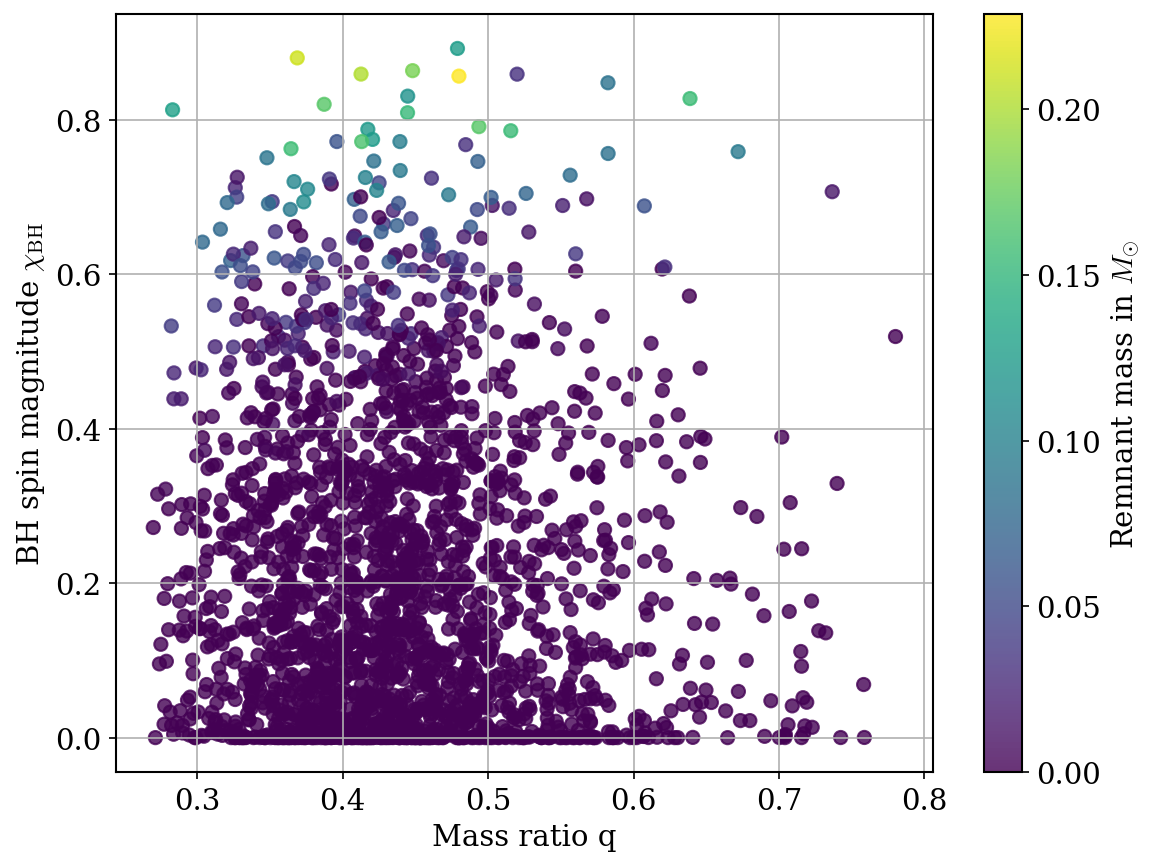}
        \caption{Pop2}
        \label{fig:rem_mass2}
    \end{subfigure}
    \caption{Mass ratio and black-hole spin magnitude for GW-detected NSBH mergers in the LHVA network for Pop1 and Pop2 (See Sec.~\ref{sec:pop_gen}). The color scale shows the remnant mass outside the black hole after the merger. Systems with lower mass ratios and higher black-hole spins are more likely to leave substantial remnant material. Pop1 produces a larger number of events with non-zero remnant mass than Pop2 owing to its broader tidal-deformability distribution.}
    \label{fig:rem_mass}
\end{figure*}

For Pop1 and Pop2, the remnant mass remaining outside the ISCO due to tidal disruption for GW-detected events in the LHVA network is shown in Fig.~\ref{fig:rem_mass} as a function of the binary mass ratio and black-hole spin magnitude. For Pop1, 55.36\% of the GW-detected sources in the LHV network have a non-zero remnant mass, corresponding to 0.53\% of the total injections. In the LHVA network, this fraction increases to 65.83\% of the GW-detected sources, corresponding to 1.16\% of the total injections. Consequently, the number of GW-detected events with a non-zero remnant mass increases by a factor of 2.19 with the inclusion of LIGO-India.

For Pop2, 11.08\% and 12.84\% of the GW-detected sources have a non-zero remnant mass in the LHV and LHVA networks, respectively. The total number of such events, therefore, increases by a factor of 2.21 with the inclusion of LIGO-India. Compared to Pop1, however, the fraction of the total injection population that is both GW detected and has a non-zero remnant mass is lower by factors of approximately 3.78 and 3.74 for the LHV and LHVA networks, respectively.

A GW-detected event with a non-zero remnant mass is a potential candidate for EM follow-up only if it can also be localized sufficiently well on the sky. To assess the feasibility of follow-up with the Rubin Observatory, we require the 90\% credible sky-localization area to be smaller than or equal to $500\,\mathrm{deg}^2$. Events with substantially larger localization regions are difficult to tile efficiently and are therefore less practical targets for EM follow-up. The choice of $500\,\mathrm{deg}^2$ provides a conservative localization criterion motivated by \citet{andreoni2024rubin2024envisioningvera}.

Applying this localization criterion to Pop1, 73.96\% of the GW-detected events with a non-zero remnant mass in the LHV network have a 90\% localization area of $\leq500\,\mathrm{deg}^2$. This fraction increases to 95.13\% for the LHVA network. Combining the improvement in the number of GW-detected events with non-zero remnant mass with the improved sky localization, the number of events suitable for EM follow-up is 2.82 times larger in the LHVA network than in the LHV network.

A similar improvement is observed for Pop2. In the LHV network, 77.95\% of GW-detected events with a non-zero remnant mass satisfy the $500\,\mathrm{deg}^2$ localization criterion, compared to 97.15\% in the LHVA network. Consequently, the number of events suitable for EM follow-up is 2.73 times larger with the LHVA network than with the LHV network, comparable to the improvement obtained for Pop1.

\begin{figure}
    \centering
    \includegraphics[width=0.96\linewidth]{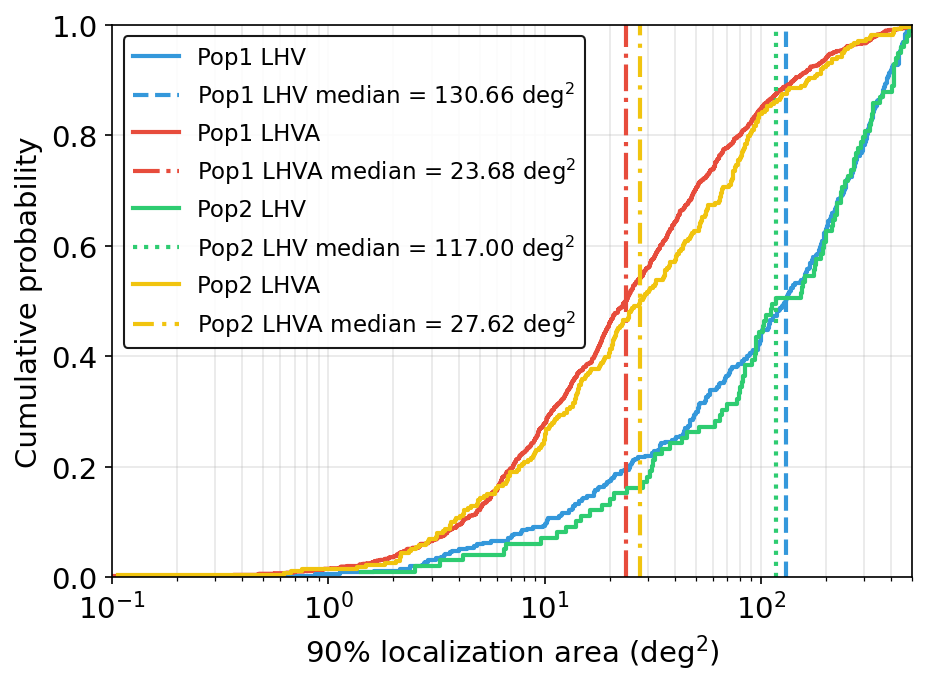}
    \caption{Cumulative probability of 90\% sky localization area for LHV and LHVA networks for detected events with non-zero remnant mass and sky localization area smaller than or equal to 500 $\mathrm{deg}^2$ for Pop1 and Pop2 (See Sec.~\ref{sec:pop_gen}). For Pop1, the median is 130.66 and 23.68 $\mathrm{deg}^2$ for LHV and LHVA, respectively. For Pop2, the median is 117.00 and 27.62 $\mathrm{deg}^2$ for LHV and LHVA, respectively.}
    \label{fig:cumarea}
\end{figure}

The distributions of the 90\% credible sky-localization areas for GW-detected events with non-zero remnant mass are shown in Fig.~\ref{fig:cumarea} for both populations and detector networks. The inclusion of LIGO-India substantially improves the localization of these events. For Pop1, the median 90\% localization area decreases from $130.66\,\mathrm{deg}^2$ with LHV to $23.68\,\mathrm{deg}^2$ with LHVA, corresponding to an improvement by a factor of 5.52. For Pop2, the median decreases from $117.00\,\mathrm{deg}^2$ to $27.62\,\mathrm{deg}^2$, an improvement by a factor of $4.24$.

\subsection{Rubin detectability and exposure-time trade-off}

\begin{figure*}[t]
    \centering

    \begin{subfigure}{0.48\textwidth}
        \centering
        \includegraphics[width=\linewidth]{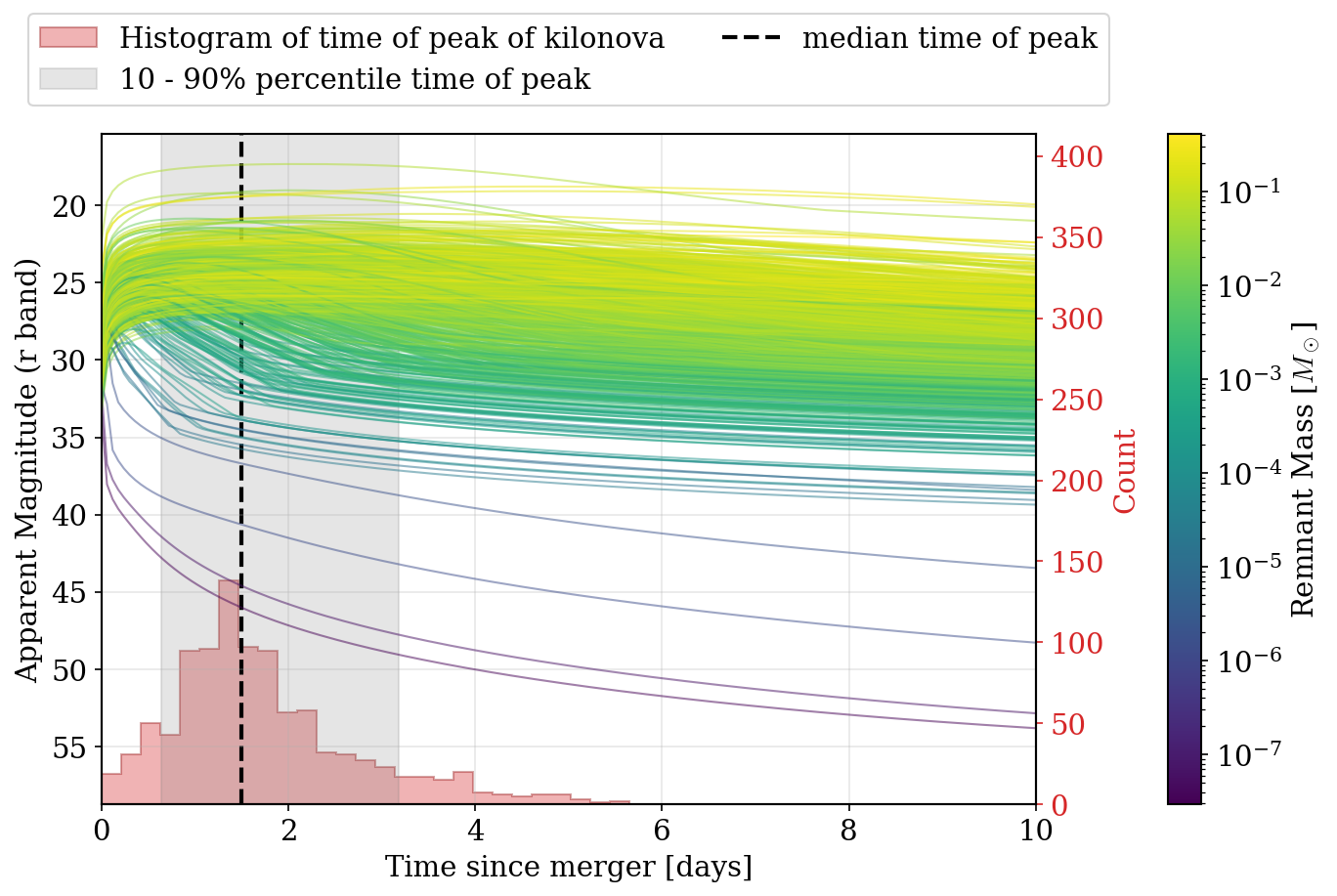}
        \caption{Pop1}
        \label{fig:lc_non_zero_5001}
    \end{subfigure}
    \hfill
    \begin{subfigure}{0.48\textwidth}
        \centering
        \includegraphics[width=\linewidth]{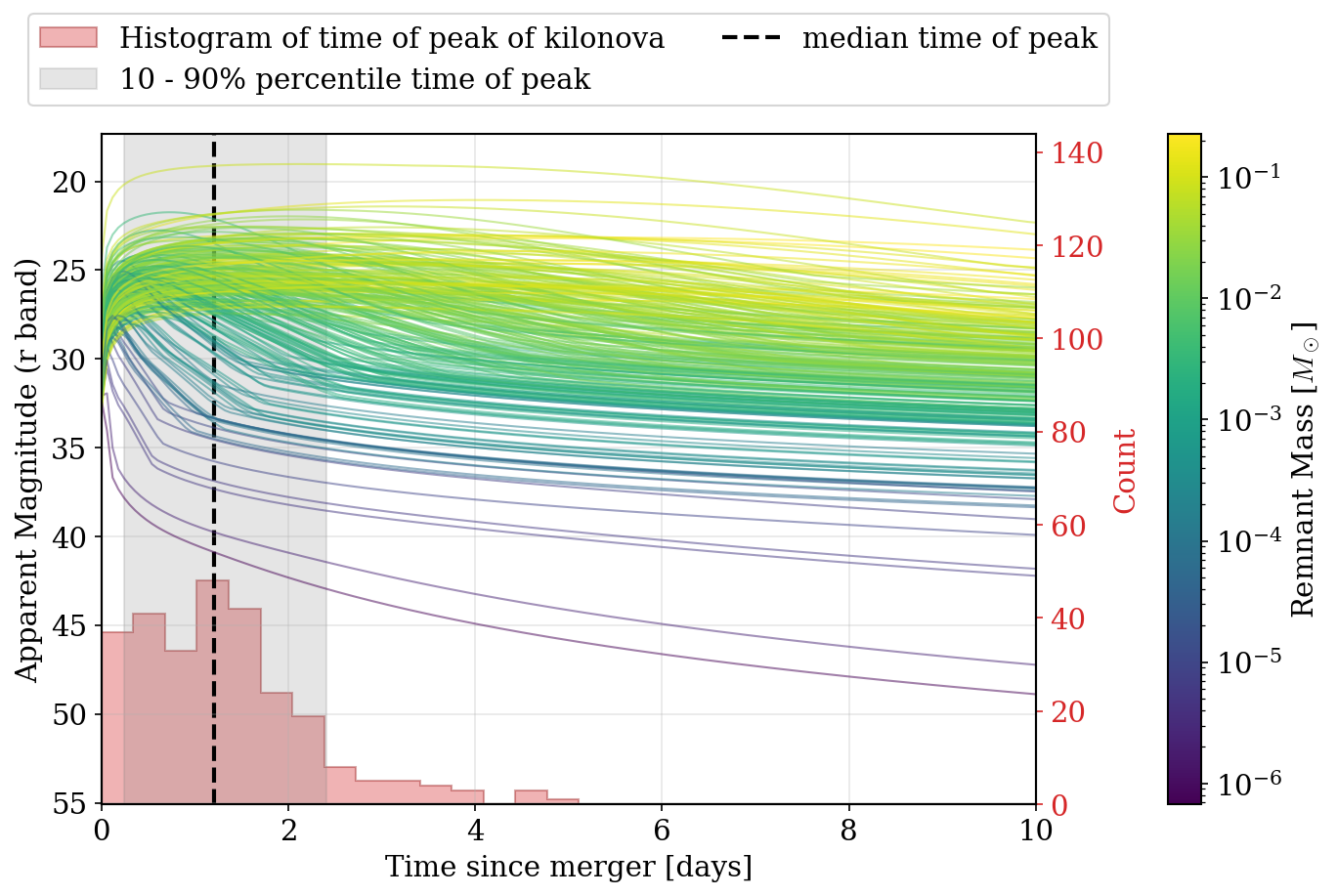}
        \caption{Pop2}
        \label{fig:lc_non_zero_5002}
    \end{subfigure}

    \caption{Lightcurves for all GW events detected in the LHVA network with non-zero remnant mass and sky localization area smaller than 500 $\mathrm{deg}^2$ for Pop1 and Pop2 (See Sec.~\ref{sec:pop_gen}). The red histogram shows the distribution of peak times in the lightcurve. For Pop1, the maximum lightcurves peak at 1.36 days, and the median time of peak(black dashed line) is 1.50 days, with 10 and 90\% events (shaded area) beyond 0.64 and 3.19 days, respectively. For Pop2, the maximum lightcurves peak at 1.19 days, and the median time of peak(black dashed line) is 1.20 days, with 10 and 90\% events (shaded area) beyond 0.24 and 2.40 days, respectively.  }
    \label{fig:lc_non_zero_500}
\end{figure*}

We use \texttt{GWEMOPT} toolkit to simulate Rubin follow-up observations of the EM candidates identified in Sec.~\ref{sec:results_gw}. For each candidate, we generated kilonova lightcurves using~\citet{gompertz_multimessenger_2023}. The resulting lightcurves for the LHVA network are shown in Fig.~\ref{fig:lc_non_zero_500}, along with the distributions of their peak times. 

For Pop1, the median time to peak is 1.50 days after the merger, with 90\% of events peaking within approximately 3.19 days. For Pop2, the median time to peak is around 1.20 days for LHVA candidates, and 90\% of events occur within 2.40 days. Similar peak-time distributions are obtained for the LHV network for both populations. These results suggest that follow-up observations should commence within about $1-2$ days of the GW trigger to capture most potential counterparts near peak brightness. Throughout this work, we therefore assume that Rubin's observations begin 1 day after the merger. 

\begin{figure}
    \centering
    \includegraphics[width=0.96\linewidth]{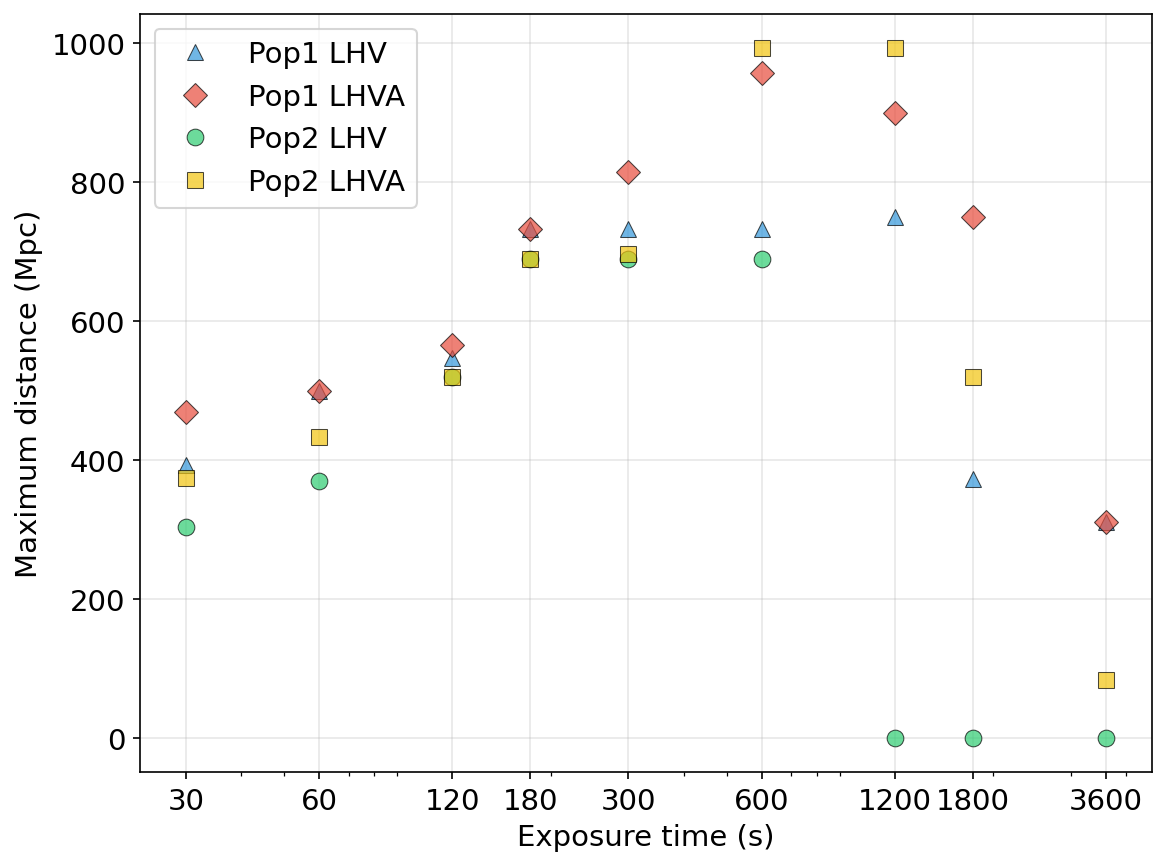}
    \caption{Maximum distance injection out of all GW detections in LHVA and LHV with successful EM follow-up in Rubin for Pop1 and Pop2 (See Sec.~\ref{sec:pop_gen}).}
    \label{fig:max_dist}
\end{figure}

\begin{figure}
    \centering
    \includegraphics[width=0.96\linewidth]{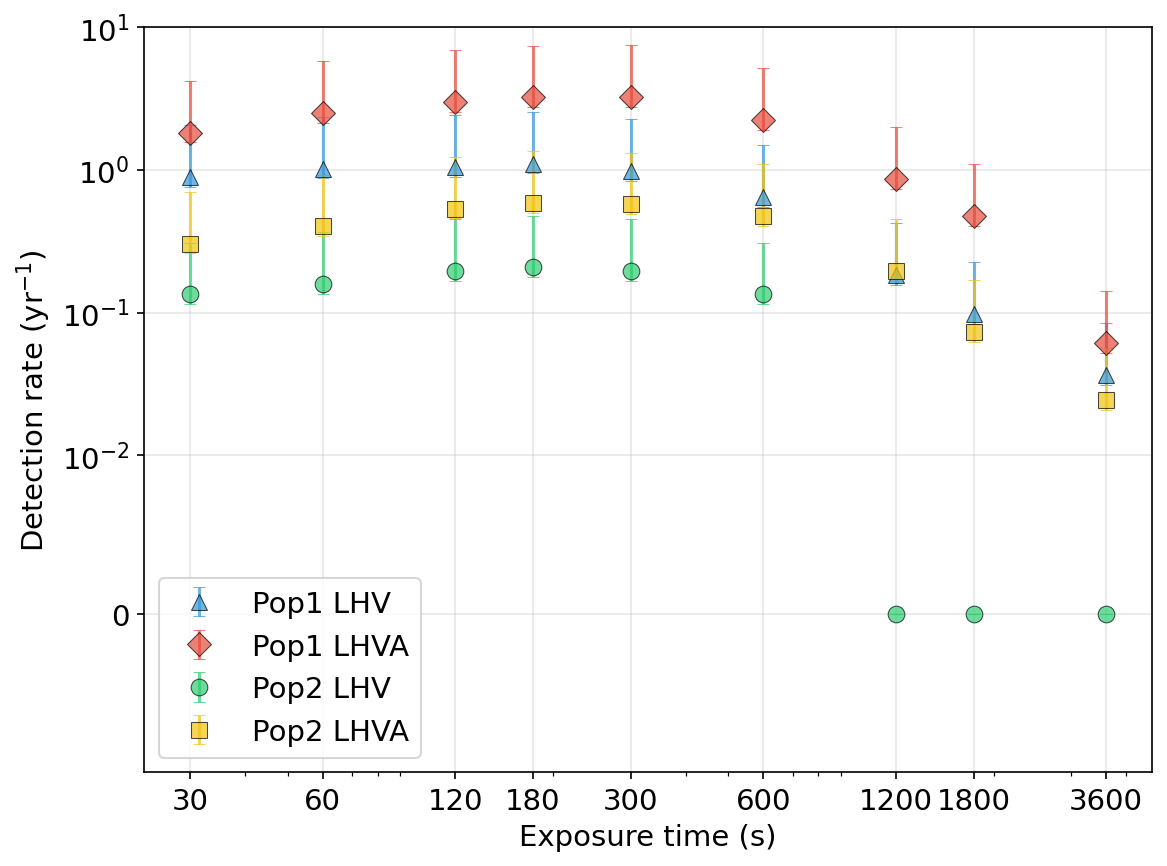}
    \caption{EM detection rate as a function of exposure time for Pop1 and Pop2 (See Sec.~\ref{sec:pop_gen}). Here, EM detections are for GW events detected in the LHV and LHVA networks with non-zero remnant mass and sky localization area smaller than 500 $\mathrm{deg}^2$.Here, the y-axis scale is logarithmic above 0.01 so that 0 can be shown on the same plot.}
    \label{fig:det_rate}
\end{figure}

The improved sky localization provided by the LHVA network also enhances the detectability of more distant counterparts. As shown in Fig.~\ref{fig:max_dist}, the maximum distance of the joint GW and EM detections increases with exposure time, and the LHVA network consistently reaches larger distances than the LHV network. To estimate realistic detection rates, we converted the number of detections obtained in our simulations into annual event rates using the merger-rate density of GW230529-like systems ($ 55^{+127}_{-47}\mathrm{Gpc}^{-3}\mathrm{yr}^{-1}$ as reported in \citet{Abac2024}). The resulting rates of GW detections with successful EM follow-up are shown in Fig.~\ref{fig:det_rate}. The detection rate increases with exposure time and reaches a maximum for exposure times of a few 100 secs. After that, longer exposure times lead to scheduling problems, resulting in fewer detections. 

The joint detection rate for the LHVA network is consistently higher than that for the LHV network, with improvements ranging from a factor of $\sim 1.7-5 $, depending on the exposure time and population model. The relative advantage of the LHVA network becomes more pronounced at longer exposure times, where efficient scheduling is increasingly limited by the sky-localization area. The improved localization provided by LIGO-India enables a larger fraction of events to be covered within the available observing time, thereby increasing the number of detectable counterparts. The total Rubin observing time required to follow up all the scheduled events is typically larger for the LHVA network by a factor of $\sim1.67 -4.88$ and $\sim1.22-2.03$ compared to the LHV network for Pop1 and Pop2, respectively, reflecting the higher number of detectable counterparts. Nevertheless, as shown in Table~\ref{table:configure} and Table~\ref{table:configure2}, the median annual observing time remains below 24 hours for all configurations considered. This is comfortably within the current Rubin ToO allocation of approximately $50-60$ hours per year~\cite{andreoni2024rubin2024envisioningvera}, indicating that the proposed follow-up strategy is operationally feasible within existing observing constraints. The relative improvement in detection rate with respect to LHV increased with an increase in exposure time, as scheduling becomes difficult for a large sky localization area in the LHV network until a point where the sky localization area becomes a factor for the LHVA network as well.

\section{optimized strategy for EM follow up}\label{sec:opt-strat}

% As we increase exposure time, a few sources with a large sky localization area are missed, and a few new ones are detected. 
The results of Sec.~\ref{sec:result} demonstrate that a single exposure time does not maximize the number of successful EM followup observations. Increasing the exposure time improves the detectability of intrinsically faint kilonovae, but simultaneously reduces the sky area that can be covered within the available observing window. Consequently, events with poor sky localization become increasingly difficult to follow up. This depth–coverage trade-off suggests that the optimal exposure time should depend on the properties of the individual GW event rather than being fixed for all observations.
%This trade-off suggests that optimal exposure time should depend on the properties of the individual event rather than being fixed for all follow-up observations.  
%This motivated us to come up with a strategy to decide the exposure time with which one can observe a particular event based on the sky localization area and apparent magnitude.

To construct such an event-specific observing strategy, we combine all successful GW+EM detections from both detector networks and both population models into a single dataset. Each detection is characterized by its peak apparent magnitude and 90\% sky-localization area. For every event, we determine the minimum exposure time that results in a successful detection. The objective is then to derive a simple prescription that assigns an exposure time based on these observable quantities while preserving the detectability of the event.

\begin{figure}
    \centering
    \includegraphics[width=0.96\linewidth]{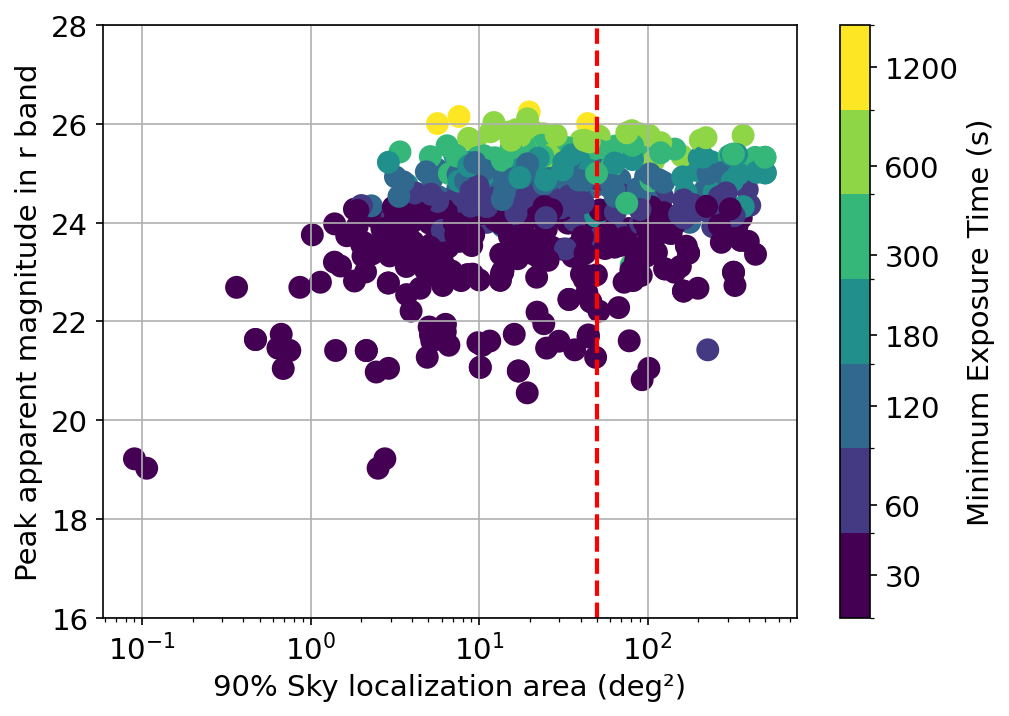}
    \caption{Peak apparent magnitude in r-band and 90\% sky localization area of all detections from in LHVA and LHV networks for Pop1 and Pop2 (See Sec.~\ref{sec:pop_gen}), with minimum exposure time of detection as colour. The red line represents a 50 $\mathrm{deg}^2$, 90\% sky localization area, beyond which there is no detection at exposure times of 1200 secs or longer. This plot shows how the minimum exposure time required for detection changes with the peak apparent magnitude and 90\% sky localization area of the source.}
    \label{fig:combine}
\end{figure}

Fig.~\ref{fig:combine} shows the distribution of successful detections, where the color indicates the minimum exposure time required for each event. It is evident from the figure that going beyond 600 secs of exposure time for sources with a sky area of more than 50 $\mathrm{deg}^2$ is unlikely to lead to any detection, as sky coverage becomes prohibitive. We restrict the exposure time for sources with a sky localization area of more than 50 $\mathrm{deg}^2$ to 600 secs. We constrain our observation time to 1200 secs for the remaining sources, as for higher exposure times, there are no new detections. From Fig.~\ref{fig:combine}, it can be seen that the required exposure time increases with an increase in peak apparent magnitude.

% We assume that if an event was detectable with any particular exposure time, then our strategy should aim to ensure that this event remains detectable in this new scheme. 

\begin{figure*}[ht!]
\centering
\includegraphics[width=0.48\textwidth]{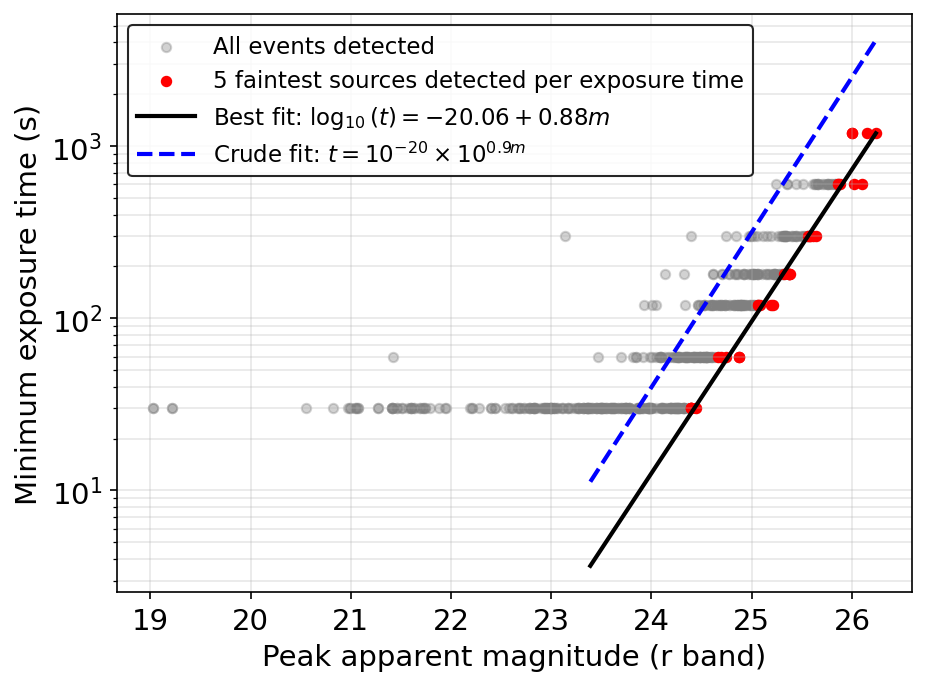}
\hfill
\includegraphics[width=0.48\textwidth]{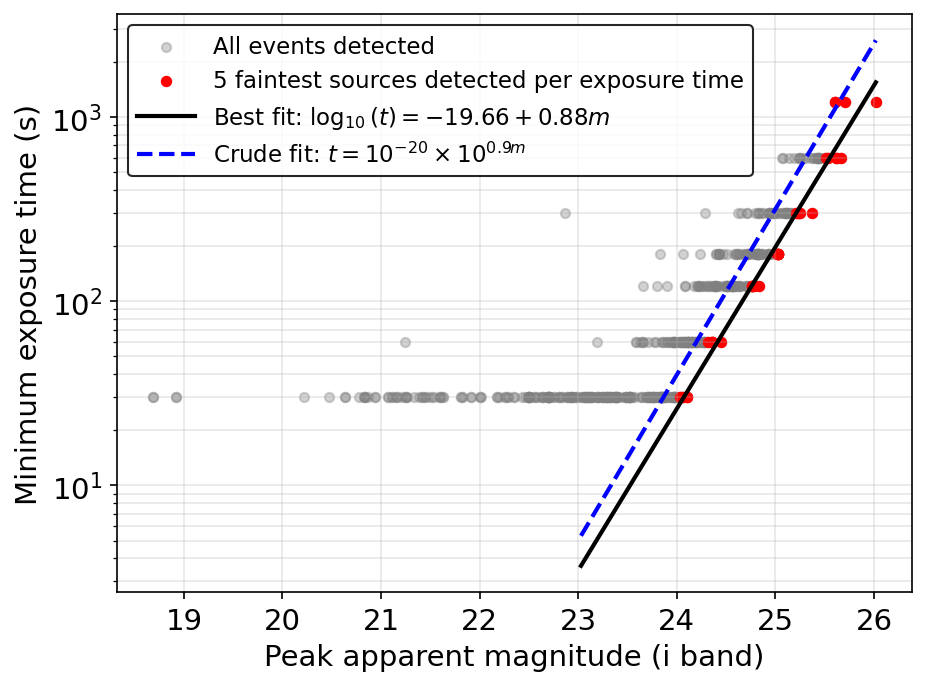}
\caption{Exposure time-- peak apparent magnitude relation in the $r$- and $i$-bands. The best-fit curves are obtained by taking the five faintest sources detected at a particular magnitude in each band. The crude fit is an approximate fit by hand.}
\label{fig:fit}
\end{figure*}

Peak apparent magnitude and minimum exposure time with which an event is detected are shown in Fig.~\ref{fig:fit}. We assume a logarithmic dependence of apparent magnitude on the exposure time required and tried to fit the 5 faintest sources detected with an exposure time. We invoke a crude fitting formula,

\begin{align}
    t_{\mathrm{exp}}=10^{-20+0.9m}\,,
\end{align}

where $t_{\mathrm{exp}}$ is the minimum exposure time required to obtain a source with a peak apparent magnitude $m$. Sources that are brighter can also be detected with that exposure time. All exposure times were rounded up to the next multiple of 30 secs.

% \begin{figure*}[ht!]
%     \centering

%     \begin{subfigure}{0.48\textwidth}
%         \centering
%         \includegraphics[width=\linewidth]{fit_r.png}
%         \caption{Peak apparent magnitude in the r-band and minimum exposure time at which the source is detected. }
%         \label{fig:fit1}
%     \end{subfigure}
%     \hfill
%     \begin{subfigure}{0.48\textwidth}
%         \centering
%         \includegraphics[width=\linewidth]{fit_i.png}
%         \caption{Peak apparent magnitude in the i-band and minimum exposure time at which the source is detected.}
%         \label{fig:fit2}
%     \end{subfigure}

%     \caption{Exposure time - apparent magnitude relation. The best fit curves are obtained by taking the 5 faintest sources detected at a particular magnitude in either band. The crude fit is an approximation fit by hand.}
%     \label{fig:fit}
% \end{figure*}

Now, for our samples, we calculate peak apparent magnitude using distance and remnant mass information (in practice, this information will become available from the matched filtering or parameter estimation results of GW-detected events). We use the source frame remnant mass ($M_{\mathrm{rem}}$) and peak absolute magnitude ($M$) relation found by fitting all the samples using the following relation:

\begin{align}
    M=-15-\log_{10}(M_{\mathrm{rem}}/M_\odot)\,.
\end{align}

\begin{table}
\begin{tabular}{c|cc|cc}
\hline\hline
\textbf{Exposure} & \multicolumn{2}{c|}{\textbf{Detection rate}} & \multicolumn{2}{c}{\textbf{Total duration}} \\
\textbf{time (sec)} & \multicolumn{2}{c|}{\textbf{(per year)}} & \multicolumn{2}{c}{\textbf{(hours per year)}} \\
\cline{2-5}
     & LHV & LHVA & LHV & LHVA \\
\hline
\rule{0pt}{3ex} Pop I & $1.46_{-1.25}^{+3.37}$& $4.19_{-3.58}^{+9.66}$& $2.65_{-2.27}^{+6.12}$& $7.37_{-6.30}^{+17.02}$ \\
\rule{0pt}{3ex} Pop II & $0.27_{-0.23}^{+0.62}$& $0.79_{-0.68}^{+1.83}$& $1.11_{-0.95}^{+2.56}$& $3.15_{-2.69}^{+7.26}$ \\
[0.15cm]
\hline\hline
\end{tabular}
\caption{EM counterpart detection rate and time required over the telescope per year with optimized strategy given in Sec.~\ref{sec:opt-strat} for Pop1 and Pop2 (See Sec.~\ref{sec:pop_gen}).}
\label{table:opt_det}
\end{table}

\begin{figure}[ht!]
    \centering
    \includegraphics[width=0.96\linewidth]{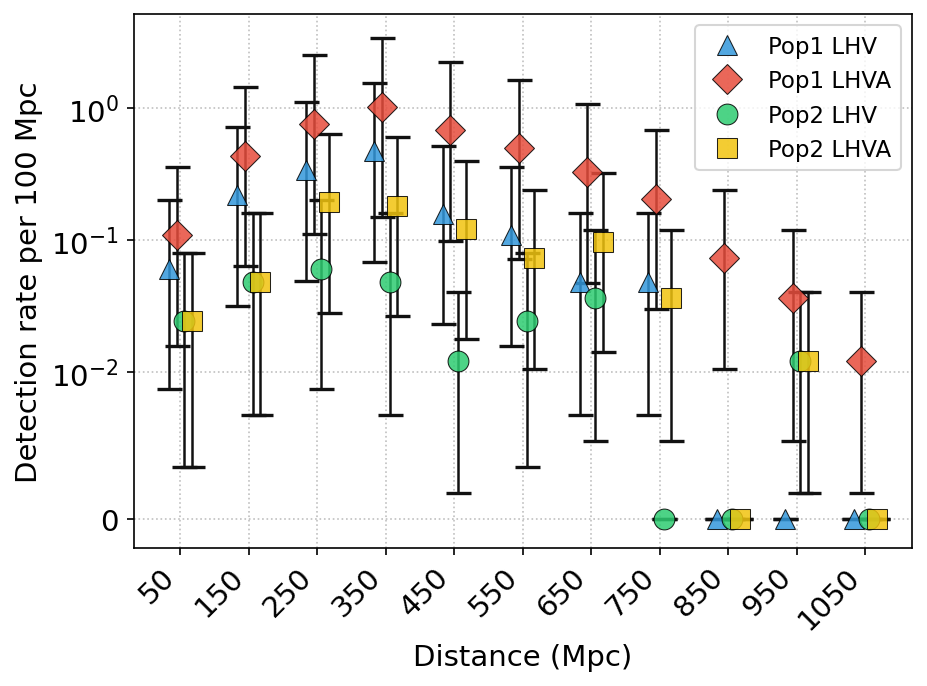}
    \caption{EM counterpart detection rate as a function of luminosity distance for Pop1 and Pop2 (See Sec.~\ref{sec:pop_gen}), using the optimized follow-up strategy described in Sec.~\ref{sec:opt-strat}. The rates are shown for GW events detected by the LHV and LHVA networks that have a nonzero remnant mass and a 90\% sky-localization area smaller than $500\,\mathrm{deg}^2$. The vertical axis is logarithmic above a rate of $0.01$ events per $100\,\mathrm{Mpc}$, while zero-rate bins are shown at the bottom of the plot. The fluctuations in the Pop2 rates, particularly at larger distances, result from the small number of detected events; for example, only 22 LHV detections are distributed over 11 distance bins.}
    % \caption{EM counterpart detection rate as a function of distance, with optimized strategy as defined in Sec.~\ref{sec:opt-strat} for Pop1 and Pop2. Here, EM detections are for GW events detected in the LHV and LHVA networks with non-zero remnant mass and sky localization area smaller than 500 $\mathrm{deg}^2$. Here, the y-axis scale is logarithmic above 0.01 so that 0 can be shown on the same plot. Fluctuations in the case of Pop2 are due to a very low number of detections (22 detections for LHV distributed over 11 bins). An increase in the number of injections can help to improve these fluctuations.}
    \label{fig:em_dist_distri}
\end{figure}

With this optimization strategy, we observe the events for up to 2 days, starting from 1 day after the merger in the r and i bands, twice each. This was able to improve the detection rate, as shown in Table~\ref{table:opt_det} as compared to the rate for any specific exposure time present in Table~\ref{table:configure} and Table~\ref{table:configure2}. The detection rate using this strategy as a function of distance is shown in Fig.~\ref{fig:em_dist_distri}. This clearly summarizes the final improvement in EM follow-up rate after introducing LIGO-India and how the distance reach is significantly improved.

%Conclude
\section{Conclusion and Discussion}\label{sec:conclusion}

Multimessenger observations of NSBH mergers, through GW and EM, though presently elusive, can lead to rich science not otherwise accessible. For instance, they can enable an independent, precise estimation of the Hubble constant and even allow probing the Hubble parameter. In this work,
%we investigate the prospect of joint GW+EM observations of NSBH mergers with the Rubin observatory in the era of LIGO India.
we show that the inclusion of LIGO-India in the ground-based detector network can substantially improve the prospects for EM follow-up by increasing the GW detection rate, improving sky localization, and increasing the network duty cycle. For the population considered here, the median expected time to one NSBH detection with successful EM follow-up by Rubin, using an optimized exposure strategy, ranges from 8.22 months to 3.74 years for the LHV network and from 2.87 months to 1.26 years for the LHVA network for populations 1 and 2, respectively. The inclusion of LIGO-India increases the expected number of GW detections from NSBH mergers by approximately a factor of two. The corresponding increase in the number of events that can be identified as EM counterparts is estimated to be a factor of $\sim 2-5$, depending on the exposure time, owing to the combined effects of the increased GW detection rate and improved sky localization. With an optimized strategy, which maximizes the total number of EM counterpart detection, the improvement factor is more than 2.9. Finally, we find that the total target-of-opportunity observing time required to follow NSBH mergers with a nonzero remnant mass and a 90\% credible sky-localization area below $500\,\mathrm{deg}^2$ is within the ToO time available at the Rubin Observatory.

% For the optimized observing strategy considered here, the expected number of joint GW–EM observations increases by more than a factor of 2.9. The time required to follow all NSBH merger events with non-zero remnant mass and a 90\% sky localization area smaller than 500 is well within the ToO time available at the Rubin Observatory.

Constraining the Hubble constant with BNS generally requires $\sim 50$ mergers; this required number can be much smaller (only a few) for NSBH systems depending on the properties of the source population~\cite{vitale_measuring_2018}. NSBH mergers with EM counterparts can provide substantially more information per event than BNS mergers in certain favorable configurations. In particular, a precessing NSBH merger with EM follow-up can provide a constraint comparable to that obtained from as many as 50 BNS mergers with EM counterparts, while a nonprecessing NSBH system can provide information comparable to that from $\sim 10$ BNS mergers~\cite{vitale_measuring_2018}. Moreover, LIGO-India can help reach farther distances, enhancing the possibility of probing the Hubble parameter in the $z \sim 0.1-0.2$ range. For broader tidal deformability (Pop1), there is a finite possibility of obtaining a single detection at distances beyond 500 Mpc in a median time of 4.8 years and 10.5 months for the LHV and LHVA networks, respectively. For narrower tidal deformability (Pop2), there is only a finite possibility of detecting beyond 500 Mpc with the inclusion of LIGO India in a median time of 4.6 years, while for LHV, this time is 13.7 years. With  broader tidal deformability (Pop1), there is a finite possibility of obtaining a single detection at distances beyond 1 Gpc in 82 years. While these timescales are long for the sensitivities we have considered here, projections for next-generation sensitivity improvements such as LIGO-A\# suggest that these timescales could be reduced by factors of a few~\cite{Asharp,Fritschel2024PostO5}.

Few more points maybe worth noting here. First, the source populations considered in this work are motivated by the possibility that GW230529 was likely an NSBH merger, although other interpretations cannot be ruled out~\cite{mali2026}. %Therefore, the actual NSBH population may differ from the models adopted here, leading to different predicted detection and EM-follow-up rates.
Second, the model in~\citet{gompertz_multimessenger_2023}  assumes blackbody radiation. This means that the generated lightcurves may not be accurate at later times as the transient evolves. Since our follow-up strategy is based primarily on the early-time lightcurve, where this model is expected to be more relevant, we do not expect this limitation to substantially affect the results presented here. We also assume that the luminosity distance and remnant mass inferred from the detected GW events are known when optimizing the EM follow-up strategy. In practice, both quantities will have measured uncertainties and should be obtained from standard Bayesian inference analysis. Propagating these uncertainties could lead to modify optimal exposure times and therefore the absolute number of detected EM counterparts. However, we expect that the relative improvement associated with the inclusion of LIGO-India to be less sensitive to these uncertainties. 

The optimized EM follow-up strategy developed here is based on the assumption that the exposure time can be determined primarily from the expected apparent magnitude of the counterpart. This provides a simple prescription for exploring the impact of different observing strategies but does not capture the complexity of scheduling ToO observations. More sophisticated approaches, potentially incorporating machine-learning methods and additional information such as the evolving lightcurve, sky-localization probability, observing conditions, and telescope scheduling constraints, could further improve the follow-up strategy.

Our estimated probability or rate of detecting EM follow-up with second-generation detectors strongly depends on assumptions about the NSBH population and its EM counterpart. However, the improvement factor that we expect with the inclusion of LIGO-India may not significantly vary across different populations. Moreover, with the ever-increasing sensitivity of the detectors, the actual rates may be well above what we have estimated, though they involves uncertainties arising from the astrophysical models.

% Recognise who helped
\section{Acknowledgement}
The authors acknowledge the use of the IUCAA LDG cluster, Sarathi, for computational work. We used software tools \texttt{PyCBC}~\cite{Usman_2016}, \texttt{MOSFiT}, and \texttt{GWEMOPT}. We would like to thank Igor Andreoni, Saurabh Magare, Anindya Ganguly, and Anupreeta More for their help with kilonova modeling and Vera C Rubin Observatory-related information. We acknowledge the discussion with Colm Talbot and Lalit Pathak regarding the NSBH merger population. We also like to thank Anirban Kopty for his help with \texttt{PyCBC} and \texttt{Bayestar}.

\section{Data Availability}

All necessary files and scripts used in this work are publicly available in our  \href{https://github.com/Yogita-Kumari/Prospects-of-EM-Follow-up-of-NSBH-mergers-in-the-LIGO-India-Era}{GitHub repository}.

\FloatBarrier % Prevents any figures from floating past this point
% \clearpage    % Forces all figures to be printed on their own page before moving on

%\clearpage
% \onecolumngrid
% --- Appendix Section ---
\appendix

\section{Order of magnitude Estimate of NSBH Detectability}
\label{sec:detect}
To get an order-of-magnitude estimate of how the number of detections changes with distance, we use a simple scenario with 100 sources, with BH and NS masses uniformly sampled in the ranges $[3,12]M {\odot}$ and $[1,2.5]M_{\odot}$, with randomly assigned inclination and polarisation. The probability of detecting the source at a particular redshift(z) depends on the antenna pattern of GW detectors. As the possibility of detection varies with direction, in order to find the probability of detection of a source at a particular redshift, the whole sky is divided uniformly into 48 directions.
%, and it is calculated whether that source is detected or not in each of these directions.
%
Each source is placed in these 48 directions. For each direction, redshift at which its optimal network SNR falls below 12 (this value of SNR threshold is taken based on \cite{singer_first_2014}) when LHVA detectors are active, is considered the redshift threshold above which that source is no longer detectable. This value is calculated by increasing redshift and calculating the corresponding SNR till the point when SNR fall below 12. Redshift exactly at 12 SNR is calculated by linear interpolation. These redshifts, at which SNR is 12, are saved as an array of increasing order. For this particular source, as redshift increases, the probability of detection will be 1 until redshift is below the first element of the array. When it crosses the first element in the array, its probability decreases by a factor of 1/48. As redshift is increased further, its probability decreases further when encountering the second element in the array and so on. For redshifts above the last element in the array, the probability of detection is taken as zero. 
In this way, we obtained a probability function of redshift, and interpolation is performed between these 48 points using a polynomial fit. 
For all sources, the probability of detection is calculated at a given $z$. 

We take a flat universe consisting of matter and dark energy with Hubble constant $H_0=70~\mathrm{km~s}^{-1}\mathrm{Mpc}^{-1}$, matter energy density parameter $\Omega_{m0}=0.3$ and dark energy density parameter $\Omega_{\Lambda}=0.7$.
To go from probability at a particular redshift to luminosity distance ($d_L$), we convert this probability $p(z)$ to $p(d_L)$ using $p(z)dz=p(d_L)dd_L$. This implies that the probability of getting a detection at that $p(d_L)=p(z)\frac{dz}{dd_L}$.The luminosity distance and redshift are related by $d_L=(1+z)d_C$, where $d_C=\frac{c}{H0}\int_{0}^z \frac{dz'}{E(z)}$ is the comoving distance where $E(z)=\sqrt{\Omega_{m0}(1+z)^3 +\Omega_{\Lambda}}~$\cite{hogg2000distancemeasurescosmology}, and,
\begin{align}
p(d_L)=p(z){\left[\frac{d_L}{1+z}+\frac{c}{H_0}\frac{(1+z)}{E(z)}\right]^{-1}} \,.
\end{align}

\begin{figure}
    \centering
    \includegraphics[width=0.96\linewidth]{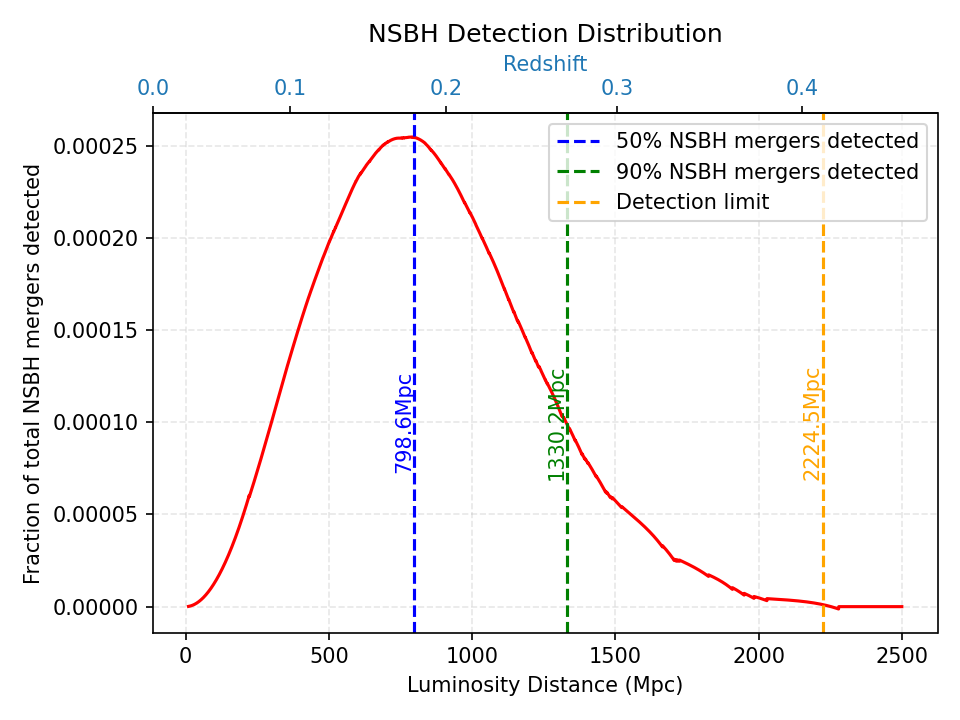}
    \caption{Order of magnitude estimate of the number of fraction of detections as a function of luminosity distance. Median distance is 798.6Mpc. 10\% detections are beyond 1330.2Mpc distance}
    \label{fig:hd}
\end{figure}

As one goes further out, detection probability decreases, but the number of sources increases, so total detected sources (integrated over mass) at a distance $d_L$ will be proportional to $d_L^2p(d_L)$. This lead to a curve as shown in Fig.~\ref{fig:hd}. From these calculations, we estimate that most detections will be within a luminosity distance of $[10,2500]$ Mpc. We distribute NSBH sources in the distance based on this in Sec.~\ref{sec:method} and that there is a non-negligible probability of detecting NSBH mergers at $\sim 1$~Gpc distances.

\bibliographystyle{apsrev4-2}
\bibliography{bib}

\end{document}